\documentclass[12pt,aasms4]{aastex}
\title {The Dependence of the Galaxy Luminosity Function on Environment}
\author {Daniel Christlein}
\affil {Steward Observatory, The University of Arizona}
\affil {933 N Cherry Ave, Tucson 85721 AZ}
\affil {and}
\affil {Max-Planck-Institut f\"ur Astrophysik}
\affil {Karl-Schwarzschild-Strasse 1, 85740 Garching, Germany}
\email {dchristlein@as.arizona.edu}
\date {\today}
\begin {document}
\begin {titlepage}
\shortauthors{Christlein}
\shorttitle{Dependence of the GLF on Environment} 
\begin {abstract}
We present luminosity functions for galaxies in loose groups in the Las Campanas Redshift Survey, differentiated by their environment (defined by the line-of-sight velocity dispersion $\sigma$ of the host groups) and also by their spectral type (emission or non-emission, defined by the equivalent width of the 3727\AA\ [OII] line).\\ We find systematic variations in the Schechter parameters $\alpha$ and $M^{*}$ for non-emission line galaxies over a range of 0 $<$ $\sigma$ $<$ 800 km/s: $\alpha$ varies from 0.20 to -0.91, indicating an increase in the steepness of the faint end slope with increasing $\sigma$. The accompanying variation in $M^{*}$ appears to be accounted for by the intrinsic correlation with $\alpha$ and does not indicate a significant physical variation in the bright end of the luminosity function. For emission line galaxies, we find no significant systematic variation of the luminosity function with the environment. Our results show that emission and non-emission galaxies generally occupy two distinct regions in the $\alpha$-$M^{*}$ parameter space. From our luminosity functions, we derive the number ratios of emission to non-emission galaxies as a function of environment and absolute magnitude, showing that the relative abundance of non-emission line galaxies generally increases for all magnitudes -23 $<$ M$_{R}$ $<$ -17.5 towards high-$\sigma$ environments, from $\sim 80\%$ to $>90\%$ at M$_{R}$ = -22 and from $\sim 10\%$ to $>50\%$ at M$_{R}$ = -18 ($H_0 = 100\:km\:s^{-1} Mpc^{-1}$ and $q_{0} = 0.5$).\\ 
\end {abstract}
\maketitle
\end {titlepage}

\section {Introduction}

The galaxy luminosity function (GLF), describing the number density of galaxies as a function of their luminosity, is a central relation in any study of galaxy evolution. For various reasons, however, global GLFs, averaged over a wide and unspecified range of spectral types, morphologies, redshifts or environments, are not the best tool imaginable for constraining such models:\\
First, hierarchical clustering models of structure formation, combined with models of galaxy formation and evolution (Diaferio et al. 1999), have the potential to predict the appearance of a group of galaxies as a function of only two input parameters besides the cosmological parameters, namely the epoch and the group mass. Clearly, a study of GLFs differentiating with respect to either of these parameters may prove more useful in constraining theoretical models than an undifferentiating one.\\  
Furthermore, if the environment is not taken into account, small surveys might suffer from sampling a non-representative region of space. For instance, the presence of massive clusters in a sample may impact the resulting GLF.\\
Thirdly, because the GLF may vary as a function of morphology, spectral type, redshift and environment, it is necessary to take into account as many of these parameters as possible to discern a real physical dependence of the GLF on the environment, for example, from a correlation arising from a real dependence on the morphology, combined with a possible morphology-environment dependence.

Considerable work has been invested over the past few years in determining GLFs for different morphologies or spectral types of galaxies (e.g. \citet{folkes}, \citet{marzke}, \citet{marzke94}, \citet{loveday}, \citet{binggeli}). Differentiation with respect to environment, however, has usually been relatively crude and only between low and high density, or field and cluster environments.\\
Here, we are presenting a systematic study of GLFs of low-redshift (cz $<$ 60,000 km/s) galaxies in groups in the Las Campanas Redshift Survey (LCRS; \citet{shectman}), differentiated by the group line-of-sight velocity dispersion $\sigma$, which allows us to examine correlations of the GLF with the environment continuously from the field up to the high-$\sigma$ environment of rich clusters ($\sigma$ $\simeq$ 800 km s$^{-1}$).

We also consider emission line (EL) and non-emission line (NEL) galaxies separately. The use of $\sigma$ as one of the parameters makes it possible to relate our results to the virial mass of the host groups, the only free parameter which the local (e.g. low-redshift) GLF should be dependent on in hierarchical clustering models. Thus, this data may prove useful in constraining models of galaxy formation and evolution in different environments (see, however, \citet{diaferio93} for a cautionary note on interpreting line-of-sight velocities as a measure of the virial mass).\\
Our consideration of two distinct spectral types in a wide variety of environments also allows us to map the $\alpha$-M$^{*}$ parameter space and the regions occupied by different types of galaxies in a systematic fashion. This approach leads us to a better evaluation of the physical significance of the correlation of M$^{*}$ with $\sigma$ which is observable in the data.

Density-dependent GLFs in the LCRS have also been presented by \citet{bromley}. Our approach differs from theirs mainly in
a different definition of the environment (galaxies in groups, selected by the group velocity dispersion, rather than galaxies selected by local density) and in a finer differentiation with respect to the environment (six subsamples selected by $\sigma$ plus the field, instead of two). Therefore, we can study the variation of the Schechter parameters with environment more systematically, while their work uses a finer differentiation with respect to the spectral type \citep{bromleyapj}.

In this paper, we will first briefly discuss the source of the data used in our analysis, namely the Las Campanas Redshift Survey, and then outline our procedure for compiling a group catalogue, preparing subsamples and analyzing the GLFs. We will then present our numerical results for the GLFs, their variations with environment and spectral type, and present the dwarf/giant ratios and EL/NEL ratios in our groups, as derived from our results.

For all of our investigations, we formally assume $H_0 = 100\:km\:s^{-1} Mpc^{-1}$ and $q_{0} = 0.5$.

\section {The data}

The Las Campanas Redshift Survey \citep{shectman} contains about 24,000 galaxy redshifts from 327 fields of an area of 1.5$^{\circ}$ x 1.5$^{\circ}$ each, arranged in six strips in the southern hemisphere. 
Redshifts range up to cz $\approx$ 100,000 km s$^{-1}$, but most galaxies in our sample are found around $\sim$ 30,000 to 40,000 km s$^{-1}$. The upper and lower apparent
magnitude limits are relatively narrow (typically from m$_{R}$ = 15.0 to 17.7 mag) and often result in only a small fraction of the total number of group members
being observed. The shallowness of the sampling creates uncertainties in the determination of any group properties, such as the radius or velocity dispersion, but
not in our calculation of the luminosity functions, for which we simply add up all galaxies in the groups meeting our selection criteria.\\
The LCRS is suitable for determining luminosity functions as the magnitude limits are well defined and the sampling fraction (e.g., the
fraction of galaxies with measured redshifts, among all galaxies meeting the selection criteria) is also well known down to M$_{R}$ = -17.5 mag. For galaxies fainter than that, the completeness of the LCRS is not well understood. We therefore cut our sample at this magnitude and consider only galaxies brighter than -17.5 mag for the luminosity functions (the group identifications, however, are done with the entire set of galaxies).\\
Photometry in the LCRS was originally obtained in the Gunn r band, but the calibration has been performed in the Kron-Cousins R band \citep{lin}.

It should be noted that for 1695 galaxies meeting the photometric limits of the LCRS, no redshifts are listed because they were too close to another object for a separate fiber to be placed on their position. This constraint might have some 
impact on the identification of groups and on some of the investigations described below. The galaxy luminosity functions may be affected as the centers of dense groups may be undersampled.\\
To determine the magnitude of this effect, we perform a separate analysis, in which these galaxies are assigned the same redshift as their closest projected neighbour. The galaxies are then linked to nearby group members, if there are any. This conservative procedure adds 751 galaxies to our group catalogue. While some of our Schechter fits did change, typically by values on the order of our 1$\sigma$ errors or less, the qualitative conclusions of this paper remain the same.

\section {The group catalogue}

We use a friends-of-friends algorithm \citep{hg,nw}, which detects overdensities in redshift space, to identify those galaxies which are likely to be part of a group. Such a group is located by selecting any galaxy as a starting point and searching for other galaxies within a predefined range of linking lengths in redshift and projected separation. If such galaxies are found, we look for their neighbours in the same way. An ensemble of at least three galaxies connected this way is considered a group.\\
To account for inhomogeneities caused by the varying magnitude limits, the linking lengths have to be roughly scaled according to the expected observable galaxy number density, were the actual distribution homogeneous. We thus choose to scale the linking lengths by the cubic root of the ratio of the expected galaxy density in a 
fiducial field to the expected density in the field to be examined; calculating these expected densities requires assuming a particular luminosity function. We use the global GLF found by Lin et al. for the LCRS. (For an excellent and deeper discussion of the techniques, including the concept of the 'fiducial field', used to construct a very similar group catalogue with the LCRS data, see \citet{tucker}.)

The observed group properties in similar group catalogues derived from friends-of-friends algorithms, like harmonic radii, velocity dispersion distribution etc. are typically dependent on the choice of the basic linking lengths (Diaferio et al. 1999). However, as we are concerned with the properties of individual galaxies and not with general group parameters, our results should be less sensitive to these correlations, with contamination of the group catalogue with field galaxies being the only issue to worry about.\\
We those choose relatively generous linking parameters of
0.90 Mpc and 700 km s$^{-1}$ in a fiducial field at cz = 30,000 km s$^{-1}$, with a 100\% sampling fraction and 
magnitude limits of 15 mag $<$ m$_{R}$ $<$ 17.7 mag, which roughly corresponds to a projected overdensity of $\delta n/n \approx 40$. To avoid implausibly large linking lengths in high-z fields, we set an upper limit of 4 for the factor by which the fiducial linking lengths could be scaled.\\

To ensure the stability of our conclusions, we also experimented with different choices of linking lengths. We discuss this effect, and that of a possible contamination of the group catalogue with field galaxies, below.

\subsection {Subsamples}

For each group in our catalogue, we calculate a line-of-sight velocity dispersion 
\begin{equation}
\sigma = (\Sigma_{N} \frac{(cz_{i}-<cz>)^{2}}{N-1})^{1/2}
\end{equation}
where N is the number of galaxies in the group.\\
All galaxies found within groups within a certain range of $\sigma$
 are added up to form one subsample. We divide the range of 0 km s$^{-1}$ $<$ $\sigma$ $<$ 1000 km s$^{-1}$ into six bins, the first five of which are spaced 100 km s$^{-1}$ apart. The sixth bin covers a wider range from 500 to 1000 km s$^{-1}$, in order to improve statistics in a very poorly populated range of $\sigma$. We gathered all galaxies not identified as group members in a seventh bin designated as the 'field' subsample; however, we should point out that, due to this procedure, this field sample is very likely to be contaminated with group galaxies which dropped below the 3-member limit due to the magnitude selection limits.

The distribution of groups among the $\sigma$ bins is subject to some uncertainty because of the errors associated with the
individual galaxy redshift measurements (up to several tens of km s$^{-1}$) and the calculation of $\sigma$ from a very limited number
of group members. This effect will be discussed further below.

 We further classify the spectral properties of these galaxies by the equivalent width (EW) of the [OII] line at $\lambda$3727 \AA, kindly provided by Ann Zabludoff. The LCRS dataset from which we extracted this information is the same as that used by \citet{bromley} and \citet{lin}. We will designate galaxies with an EW $\geq$ 5 \AA\ as emission line (EL) galaxies, and those with an EW $<$ 5 \AA\ as non-emission line (NEL) galaxies.\\
In all our diagrams, data points referring to NEL galaxies will be drawn as crosses, and data points referring to EL galaxies as squares.

\subsection {The Luminosity functions}

To describe the galaxy luminosity functions within our subsamples we use the Schechter function \citep{schechter}:
\begin{eqnarray}
\Phi(M) = (0.4\:ln\:10)\:\Phi^{*}\:[10^{0.4(M^{*}-M)}]^{1+\alpha}\:exp(-10^{0.4(M^{*}-M)}) 
\end{eqnarray}
The function is characterized by two parameters: the faint end slope $\alpha$ (a more negative $\alpha$ signifying a more steeply rising slope towards the faint end) and M*, which roughly describes the magnitude at which the exponential cut-off at the bright end begins to dominate over the power law describing the faint end.

To compute the luminosity functions, we use a parametric maximum likelihood method, based on the approach by Sandage, Tamman and Yahil (1979). Independently of these calculations, we also use the non-parametric stepwise maximum likelihood method by Efstathiou, Ellis and Peterson (1988) to calculate discrete luminosity functions in bins of 0.5 mag to verify the plausibility of the Schechter fits by visual inspection. Many of the diagrams that we present in this paper show the results from both methods for comparison: those derived from the parametric Schechter fits as continuous lines and those based on the discrete GLFs as data points with error bars. It should be pointed out that these two sets of results are calculated independently of each other, and that our parametric luminosity functions and all relations derived from them are not computed as a direct fit to the discrete data points.

Both methods of calculating luminosity functions, with special reference to the LCRS, are also detailed in Lin et al. (1996). Lin et al. employ various corrections to account for questions of survey completeness that typically affect the Schechter parameters in the second decimal digit. These corrections do not significantly affect our results, so we do not use them for the calculation of $\alpha$ and M$^{*}$. However, we do take the sampling fractions given in the LCRS into account for the normalization of our GLFs. Also, in contrast to Lin et al., we normalize our luminosity functions to galaxies per group and per magnitude interval instead of number density per mangitude interval, although the distinction does not affect our conclusions. The normalization parameters that we obtain are given for reasons of completeness only, and we thus do not calculate their uncertainties.

\subsection {The mock catalogue}

To rule out sources of systematic errors anywhere along our procedure, we construct a mock group and galaxy catalogue and apply several of our algorithms to it in the same way as to the original survey. The mock catalogue is based on the null hypothesis that the luminosity functions for EL and NEL galaxies are independent of environment. Further assumptions are that the groups are distributed homogeneously in comoving space and that group sizes (e.g., the number of galaxies within a given magnitude range), as well as the ratio of EL to NEL galaxies, may vary as a function of $\sigma$. 

To construct a mock group, we first select a random position in redshift space in any of the survey fields and at any redshift up to 100,000 km s$^{-1}$, the redshift range effectively covered by the LCRS, although all of our original groups were detected below 60,000 km s$^{-1}$. We then select one group from the original survey at random to serve as a model for the new mock group. From our spectral-type-independent GLFs, we then select the GLF appropriate for the $\sigma$ of our model group and use it to extrapolate the total number of galaxies expected in that group within a standard range of absolute magnitudes from -25 mag to -15 mag. If the ratio of extrapolated to observed galaxies exceeds a factor of 4, however, we reject the mock group and select a new one, in order to avoid magnifying Poisson errors in the galaxy counts too much.

We assign the mock group the original $\sigma$ value and the same size (e.g. the same number of galaxies within the standardized magnitude range from -25 mag to -15 mag) as the model group from the original catalog. We also assign the mock group an EL/NEL ratio appropriate for a group of this $\sigma$, derived from a linear fit to the EL/NEL ratios observed in the original groups (this linear fit was actually derived via two independent methods, but both showed agreement within 10\%).
 A random number algorithm then draws a population of EL and NEL galaxies from the assumed environment-independent GLFs for EL and NEL galaxies (which we took to be Lin et al.'s 1996 GLFs).

Using the original magnitude limits and field sampling fractions from the LCRS for the selected plate, we then decide which of the mock group's galaxies would have been observed in the survey. Counting these, we then judge whether the group itself would have been part of the group catalogue by the same criteria used in the original catalogue. If so, the observed mock galaxies are added to our mock catalogue.

Our mock catalogue is designed to detect any potential systematic errors. We largely eliminate statistical errors by generating ten times more groups in the mock catalogue than were found in the original survey.

The resulting mean number of galaxies per group is consistent with that found in the original catalogue. An exact match of the two catalogues is not required, however, as the mock catalogue is merely to serve as a check on the presence or absence of systematic errors introduced by our procedure of group selection and data analysis.

\section{Results}
\subsection{Luminosity functions}

Table 1 shows the best fit $\alpha$ and M$^{*}$ values for seven samples of EL and NEL galaxies in groups of different velocity dispersions and in the field. All GLFs with their Schechter fits are displayed in Figs. 1 and 2 and are normalized to galaxies per group and per dM$_{R}$ = 1 mag magnitude interval. The discrete GLFs and the Schechter fits show reasonable agreement for most luminosities.\\
Fig. 3 shows the Schechter parameters in the M$^{*}$- $\alpha$ parameter space. The respective values for the field population are plotted with error bars for better distinction. The error ellipses for the group subsamples correspond to the 1$\sigma$ uncertainties. Note that the data points for the NEL subsamples form a well-defined sequence from the lower left to upper right in order of increasing $\sigma$. It is obvious that the EL and NEL subsamples occupy two very different regions in parameter space. No significant systematic progression of the Schechter parameters with environment is suggested for the EL subsamples, however.

\subsection {Variation of $\alpha$}

We find a strong $\sigma$-dependence of the Schechter parameter $\alpha$, and also distinct differences between the EL and NEL
subsamples, as shown in Fig. 4. Table 2 gives the slopes, significances and Spearman rank correlation coefficients obtained by fitting linear relations.

We can see that $\alpha$ appears highly negatively correlated with $\sigma$ in the NEL subsample. There is no significant trend in $\alpha$ for the EL subsamples, and even considering the uncertainties suggested by the scattering of the data points, any possible correlation of $\alpha$ and $\sigma$ must be significantly weaker than for the NEL galaxies.

Generally, $\alpha$ is more negative (the faint end slope is steeper) for EL than for NEL galaxies, particularly in low-$\sigma$ environments, but the discrepancy is markedly smaller in high-$\sigma$ regions.

\subsection {Dwarf-to-giant ratio}

A different way of presenting the variation of the faint end with $\sigma$ is to plot the ratios of numbers of dwarfs to giants. We follow \citet{zabludoff} in defining galaxies with M$_{R}$ $<$ -19.6 mag as giants and -19.6 mag $<$ M$_{R}$ $<$ -17.6 mag as dwarfs and integrate our luminosity functions to derive the dwarf-to-giant ratio (Fig. 5). The numerical values for linear fits are given in Table 3.

Reflecting the same trends as $\alpha$, the dwarf-to-giant ratio is generally higher for the EL than for the NEL subsample; for the NEL galaxies, it shows a positive correlation with $\sigma$, apparent both in the discrete and the parametric luminosity functions. There exists a discrepancy between the slopes derived from the parametric and discrete values, but Fig. 5 shows the discrepancy to be mainly due to the last data point at 750 km s$^{-1}$. While the magnitude of the slope may thus be unclear, there is clearly a significant trend (at the 14 or 11 $\sigma$ level, respectively, based on the dispersion of data points around the linear fit) in either case.\\
The slope for the EL subsample is almost as steep as that for the NEL subsample; however, the large uncertainties in this slope do not permit a conclusion as to whether there is an underlying trend or not; the weak $\alpha$-$\sigma$ correlation for EL subsamples would lead us to expect little systematic variation in the dwarf/giant ratio as well. Indeed, Fig. 5 does indicate considerable variance of $\alpha$ among the individual $\sigma$ bins, but no clear trend.

In the case of the discrete GLFs, there are no galaxies in some of the magnitude bins at the extreme bright and faint ends; in order to make these GLFs available for an evaluation of the dwarf/giant ratio, values for these bins are extrapolated linearly from the two most nearby bins that contain data.\\
We also test the effect of extrapolating the missing bins from our parametric GLFs instead. It turns out, however, that this procedure makes no significant difference in the resulting slope of the correlation of the dwarf/giant ratio with $\sigma$, because the values of the GLF in these bins are either too small (at the bright end) or outside the range of our definition of dwarfs (at the faint end) to affect the results in a significant way.

Qualitatively, a positive correlation of the dwarf-to-giant ratio with $\sigma$ is in agreement with the observations by 
Zabludoff \& Mulchaey, who find that the dwarf/giant ratio is correlated with x-ray brightness and local galaxy density.

\subsection {Variation of M$^{*}$}

Roughly indicating the transition point from the bright to the faint end, M$^{*}$ is the second parameter of the Schechter form of the luminosity function. For the NEL subsamples, M$^{*}$ is also correlated with $\sigma$, in the sense that high-$\sigma$ groups show a more negative M$^{*}$ (Table 1 and Figure 3). There is no significant trend within the EL subsamples.
However, because $\alpha$ and M$^{*}$ are not independent of each other (from visual inspection, the correlation of $\alpha$ with M$^{*}$ appears even stronger than that of either parameter with $\sigma$), the M$^{*}$-$\sigma$ correlation does not necessarily indicate a true variation in the bright end; rather, the variation in $\alpha$ might automatically lead to the best Schechter fit being found at a different M*.\\
We will therefore postpone the discussion of the physical significance of this variation to the next section.

\subsection {NEL fraction}

One of the most striking differences between galaxy populations in the field and in clusters is the higher fraction of non-emission line galaxies in the high-density environments. A detailed study of the correlation of star formation and morphological properties with environment in the LCRS can be found in \citet{hashimoto} and \citet{hashimotothesis}.

 The luminosity functions determined in the preceding sections allow us to compute the fraction of NEL galaxies in each subsample both as a function of $\sigma$ and the galaxy luminosity M$_{R}$. Fig. 6 shows this 
fraction as a function of M$_{R}$ in each of our six $\sigma$ bins. Again, continuous lines are derived from the parametric GLFs, data points from the discrete GLFs.\\
The most obvious result is that the bright end (around -22 mag) is entirely dominated by NEL galaxies for all $\sigma$; the NEL fraction is generally $>80\%$ in this region and appears to increase even more towards high-$\sigma$ systems.
For even higher luminosities, the GLFs are ill-defined; extrapolation of our
Schechter fits would suggest a lower NEL fraction at the extreme bright end for the low-$\sigma$ groups, but this is not
supported by the discrete GLFs.\\
The fraction of NEL galaxies clearly decreases towards the faint end, which is dominated by EL galaxies in low-$\sigma$ groups. The NEL fraction at the faint end is distinctly higher (with $\sim 50\%$) in high-$\sigma$ systems, however.
 
We also note that, at the faint end, the NEL fraction is generally higher (though not significantly in most subsamples) if derived from the discrete GLFs. This may suggest that the Schechter function may not be a good approximation of the observed luminosity distribution. Indeed, Fig. 2 confirms an excess of NEL galaxies over the Schechter fits for M$_{R}>$-18 mag in most subsamples. But it is also possible that the incompleteness in the LCRS at M$_{R}$ $<$ -18 mag is responsible.

\section {Discussion}

\subsection {Effects of linking lengths}

As the magnitude limits of the LCRS are relatively narrow, the local galaxy density is not very well defined, and the group catalogue is thus sensitive to the choice of linking parameters. Low values may split up larger groups, while high values may lead to a greater contamination with field galaxies.\\
It is conceivable that the fraction of interlopers and spurious groups in our catalogue is correlated with $\sigma$. If misidentifications of group members were more likely in high-$\sigma$ groups, however, we would expect the luminosity functions to approach that of the field population in the absence of a physical variation in the Schechter parameters. This is obviously not the case; our NEL luminosity functions clearly evolve away from the field population as we go towards higher $\sigma$.\\
On the other hand, if the low-$\sigma$ end is more severely contaminated with field galaxies, the shallow faint end slopes obtained for low $\sigma$ may be the consequence of including a larger fraction of field galaxies (drawn from a shallow-$\alpha$ population), and the $\alpha$-$\sigma$ correlation may not represent a true variation of the GLF in groups. Although the observed trend would still represent a variation between high and low density environments, it would not allow us to constrain the variation from low to high $\sigma$ virialized systems.\\
Indeed, data from N-body simulations (A. Diaferio, 2000, private communication) suggests that the fraction of interlopers in groups detected via a friends-of-friends algorithm is inversely proportional to $\sigma$.

Does contamination by interlopers seriously affect our conclusions? If contamination with interlopers and spurious groups were a serious problem for our catalogue, we would expect our results to change if we remove those groups from our catalogue whose identification is the most unreliable, e.g. those with a small number of observed members. However, the $\alpha$-$\sigma$ relations for NEL galaxies that we obtain from a group catalogue after removing all triplets and subsequently all quadruplets are nearly identical to that obtained from the full catalogue (Fig. 7). Thus, reducing the catalogue to a more reliable subsample does not weaken our observed trend in $\alpha$. We may then conclude that contamination does not affect the luminosity functions significantly, either because there is little contamination or because the field GLF is very similar to the low $\sigma$ group GLF.

To further estimate the systematic effects of different linking lengths on our results, we experiment with various parameter sets. Our standard choice of a projected separation of 0.9 Mpc and a redshift separation of 700 km s$^{-1}$ in the fiducial field corresponds to an overdensity of $\delta n / n \approx 40$. Additionally, we construct catalogues with linking lengths of 0.715 Mpc / 500 km s$^{-1}$ ($\delta n / n \approx 80$, the value adopted by \citet{tucker}) and 1.45 Mpc / 700 km s$^{-1}$ ($\delta n / n \approx 10$) and analyze some of their properties. \\
In all three cases, trends such as the $\alpha$-$\sigma$ relation for NEL galaxies are apparent. In particular, the $\alpha$-$\sigma$ relation has essentially the same slope for all three choices of linking lengths, but the curve is shifted on the order of $\Delta\alpha$ $\sim$ 0.2 in the positive direction from our standard choice for the larger linking lengths, and in the negative direction for smaller linking lengths (Figure 8). The most obvious explanation is that the fraction of interlopers, drawn from a field population with a very shallow faint end slope, varies with different linking lengths, but in terms of $\alpha$, the magnitude of this variation is not dependent on $\sigma$.

It is notable that some trends are even present in the very generous choice of 1.45 Mpc. This result may suggest that the properties of galaxies are correlated over scales even larger than those commonly used for identifying groups with a friends-of-friends algorithm.

\subsection {The faint end}

Strong variations at the faint end for specific spectral types, such as found by us, may be of particular interest for constraining environment-dependent mechanisms of galaxy formation and evolution, such as biased galaxy formation, ram pressure stripping or merging.\\
Our NEL subsamples clearly show a significant negative correlation for the faint end slope $\alpha$ with $\sigma$, meaning that the faint end slope of the luminosity function is steeper in high-$\sigma$ environments. The EL subsamples show considerable variance, but no significant trend in $\alpha$ vs. $\sigma$.\\
We also note that the NEL field population is found to be more similar to low- than to high-$\sigma$ groups (cautioning again that the field subsample is likely to be contaminated with group galaxies).\\
The change in $\alpha$ is not merely a result of the morphology-density relation; in other words, it cannot be explained by the varying ratio of two galaxy populations with individual luminosity functions. NEL galaxies typically have a shallower faint end slope than EL galaxies; the morphology-density relation, which states that the fraction of early-type, presumably NEL, galaxies increases towards high-$\sigma$ environments, would thus lead one to expect a positive correlation of $\alpha$ with $\sigma$, and not the observed negative one.

To further confirm this observation and rule out any sources of systematic errors in the process, we perform the same analysis on a mock catalogue. The resulting slopes are given in Table 4.

The significant positive correlation for the complete mock sample is the effect expected from the morphology-density relation which we described above. The EL and NEL subsamples within the mock catalogue, however, clearly show no significant internal variation of $\alpha$ with $\sigma$. We may thus conclude that the correlation observed in the LCRS data is real.

The dwarf/giant ratio confirms the picture obtained from our investigation of $\alpha$. There is a significant trend for the NEL galaxies, while there is a large scatter in the EL subsamples, but no significant correlation. The mock catalogue (Table 5) confirms that no systematic errors in our procedures cause this effect.

The mock catalogue also demonstrates a good agreement between the parametric and discrete values. The reason for the discrepancies observed in the LCRS data thus is not due to any systematic error introduced by our method.

The significant differences between the mock catalogue and the observational results clearly indicate that the relative abundance of dwarf NEL galaxies increases towards higher-$\sigma$ environments, while the shape of the luminosity function of EL galaxies shows little, if any, significant systematic variation with $\sigma$.

It is not obvious from our data, however, if environmental effects physically tend to increase the number of NEL dwarfs (for example, by tidal stripping) in high $\sigma$ systems, or rather lower the number of NEL giants (e.g. by merging, although one would expect merging activity to play a smaller role in high $\sigma$ environments due to the larger merging cross section at lower relative velocities). In this context, we note that the field population appears to be more similar to the low $\sigma$ than the high $\sigma$ subsamples, indicating that any environmental effects possibly related to the observed change in the GLF may play a more prominent role in the high than in the low $\sigma$ subsamples, if the field population is taken as a relatively undisturbed sample.

\subsection {The bright end}

Like $\alpha$, M* appears to be highly correlated with $\sigma$, and this correlation is only found in the NEL, but not the EL, subsample.\\
This correlation is not indicative of a physical variation of the bright end, however. The two Schechter parameters are not independent (note that the error axes in M*-$\alpha$ parameter space are not parallel to the axes of the coordinate system); consequently, a variation in $\alpha$ will also shift the best fit value of M$^{*}$ and vice versa.\\
Nevertheless, $\alpha$ is primarily descriptive of the faint end slope, and M$^{*}$ of the extension of the luminosity function towards the bright end. The variation with $\sigma$ exhibited by our Schechter fits with two degrees of freedom does not permit us to separate the variation of one particular Schechter parameter into contributions from real physical variations and from its dependence on the other parameter.\\
For this reason, we decided to reduce the degrees of freedom for the Schechter fits to one, letting only one parameter find its best fit value, while fixing the other one at a uniform value for all $\sigma$. By repeating this procedure for different uniform values, we obtain tracks in the M*-$\alpha$ parameter space, each of which is assigned to one $\sigma$ subsample. These tracks essentially show the best fit value the free parameter attains for each $\sigma$ sample if the other one is fixed at a given value. As we can either choose $\alpha$ or M$^{*}$ to be the free parameter, we obtain two sets of tracks.\\
This procedure allows us to compare the $\alpha$ values obtained for each subsample for a unified M$^{*}$, and vice versa. Of course, quantitatively the resulting variations with $\sigma$ will have little physical meaning, but their presence or absence will indicate if the contributions to the variations in one parameter are entirely due to the interdependence between M$^{*}$ and $\alpha$, or partly indicative of a physical variation of the part of the luminosity function described by that parameter.\\
This approach is also an additional step towards mapping M$^{*}$-$\alpha$ parameter space, by telling us how galaxy subsamples align if certain characteristics of their luminosity distribution vary independently.

We performed this analysis on the NEL subsample only, as it is the one showing the most obvious variations.

Fig. 9 shows the best fit M$^{*}$ for any given $\alpha$. It is obvious that the tracks essentially all lie within the error ellipses associated with the individual data points. No significant correlation remains; at $\alpha$=0.4, the Spearman coefficient is r$_{S}$=0.4, and the significance of the M*-$\sigma$ slope is only 1.35$\sigma$. In other words, if $\alpha$ does not vary, M$^{*}$ does not independently vary in our sample either.\\
We can read Fig. 9 in yet another way. If we assumed the bright end of the GLF to be unchanged for all $\sigma$, would a variation in $\alpha$ alone be able to account for the range in M$^{*}$ that is observed? Fig. 9 confirms that it would. Any one $\sigma$ subsample covers the entire range of M$^{*}$ that we obtained earlier by setting $\alpha$ to different values.\\ 
We performed the countercheck by repeating the procedure with M$^{*}$ held constant and $\alpha$ being allowed to find its best fit value. As Fig. 10 shows, the tracks in parameter space do not coincide within the error ellipses, showing that, for any given M$^{*}$, each subsample yields a different $\alpha$. Here, r$_{S}$=-0.94, and the slope is significant at a level of over 8$\sigma$ (for M$^{*}$=-20.4 mag). Of course, for a given M$^{*}$, the range covered in $\alpha$ is much smaller than in a Schechter fit with two degrees of freedom; it is this smaller variation measured for a fixed M$^{*}$ that is physically indicative of a true variation of the faint end slope of the GLF.

We thus may conclude that the observed variation of M$^{*}$ within the NEL sample is apparently accounted for by its intrinsic correlation with $\alpha$. The variation of $\alpha$, on the other hand, remains significant even if M$^{*}$ is held constant. The real physical variation occurring within the NEL subsample from low to high $\sigma$ thus seems to be the increasing slope of the faint end, while there is no intrinsic variation of the bright end as described by M$^{*}$.

Fig. 9 allows us to compare the bright end properties of EL and NEL subsamples as well, something that cannot be done meaningfully by comparing their respective M$^{*}$ alone. Although there is no deviation among different $\sigma$ subsamples of the same spectral classification, a comparison of Figs. 3 and 9 shows that there clearly is a deviation from the M$^{*}$($\alpha$) tracks in Fig. 9 as we move between different spectral subsamples. More specifically, in Fig. 3, the EL subsamples are located towards more negative $\alpha$ / fainter M$^{*}$, compared to the NEL subsamples.\\ 
The fact that EL and NEL subsamples are not aligned along common M$^{*}$($\alpha$) tracks indicate that the bright end properties of the EL and NEL subsamples are indeed different. It is in this context that we can think of NEL galaxies as being intrinsically fainter than EL galaxies, a conclusion which can only be drawn from a simultaneous comparison of $\alpha$ and M$^{*}$.

\subsection {Comparison to Bromley et al.}

\citet{bromley} use a finer differentiation with regard to spectral type, working with 6 subsamples instead of our EL/NEL distinction. However, with respect to the environment, they use a cruder differentiation, distinguishing only between high and low galaxy density environments in redshift space.\\
Our selection of group galaxies by the internal velocity dispersion of the groups carries the advantage of being more immediately connected to the group virial mass as the free parameter in hierarchical clustering models of galaxy evolution.\\
Bromley et al. find a strong variation of $\alpha$ as a function of their spectral type, ranging from $\sim$ +1 for very early-type objects to $\sim$ -2 for extreme late-type galaxies.\\
This result is generally confirmed by our observation of $\alpha$ being more negative for EL than NEL galaxies, at least in low $\sigma$ regions.\\
With regard to the environment, they find ``a significant shift in both M* and $\alpha$ for some objects''.\\ 
Our subsamples suggest that these variations occur only in what we have defined as NEL galaxies. Furthermore, our finer differentiation with respect to the environment allows us to conclude that the shift in M$^{*}$ reported by Bromley et al. is accounted for by the direct correlation of the two Schechter parameters and is not indicative of a significant physical variation of the bright end.

\section {Summary}

We have found significant intrinsic variations of the GLF in galaxy groups with environment, which we characterize by line-of-sight velocity dispersion.
In particular, there is a highly significant negative correlation with $\sigma$ of the faint end slope $\alpha$ of the GLF of NEL galaxies, suggesting that the ratio of dwarf to giant galaxies is continuously increasing from low to high mass groups. Over the range of 0 km s$^{-1}$ $<$ $\sigma$ $<$ 800 km s$^{-1}$, $\alpha$ varies with an average slope of d$\alpha$/d$\sigma$ = -0.161 $(100 km s^{-1})^{-1}$. 

Our analysis of a mock catalogue indicates that this trend is not the result of systematic errors in our procedure. Removing the most unreliable groups from our catalogue does not significantly change the observed trend, indicating that the variation at the faint end does occur within galaxy groups and is not just an artifact of including a larger fraction of field galaxies at the low $\sigma$ end. The conclusions also hold for larger and smaller sets of linking lengths.\\
The GLF of the EL subsample, on the other hand, exhibits little or no significant systematic variation with $\sigma$, albeit there is considerable scatter in the relation.

The bright end Schechter parameter M$^{*}$ is highly correlated with $\alpha$ among different $\sigma$ subsamples of the same spectral type, but beyond this correlation there appears to be no variation of M$^{*}$ with $\sigma$ that could be attributed to an independent, physical variation of the bright end with $\sigma$. Thus, there is little variation at the bright end of the GLF from low to high $\sigma$ environments.

M$^{*}$, on the the other hand, is intrinsically fainter for EL than for NEL galaxies. This is most obvious from our plot of the best fit Schechter parameters in $\alpha$-M$^{*}$ parameter space, in which EL and NEL galaxies occupy two very different regions. NEL galaxies in groups are drawn from populations with ($\alpha$;M$^{*}$) values from a narrow band running from faint M$^{*}$ and positive $\alpha$($\sim$ +0.2) for low $\sigma$ environments to bright M$^{*}$ and strongly negative $\alpha$ ($\sim$ -0.9). EL galaxies are drawn from a population located more towards faint M$^{*}$ and strongly negative $\alpha$ in $\alpha$-M$^{*}$ parameter space. The EL population does not appear to shift systematically between different $\sigma$ environments.

Our mapping of $\alpha$-M$^{*}$ parameter space may also be a useful tool in evaluating discrepancies between different independent determinations of the GLF. As we discussed in the introduction, results of surveys of the GLF may vary, depending on what typical environment has been surveyed. 
We may think of each data point in $\alpha$-M$^{*}$ parameter space as representing one statistical parent population, associated with a particular environment and spectral type, which galaxies have been drawn from. If the environment is indeed the only free parameter the parent population is correlated with, then by covering a very wide range of possible environments, our GLFs also survey a wide range of possible parent populations in $\alpha$-M$^{*}$. Even if the typical environment or the evolutionary history of a given sample of normal galaxies is not known a priori, we might thus expect their GLF to fall into one of the regions indicated by our samples. Using our mapping of $\alpha$-M$^{*}$ space, one may be able to judge whether discrepancies between different samples may be accounted for by differences in the environment in which these galaxies reside.

\acknowledgments

The author would like to express his gratitude to Simon White, for the hospitality he experienced at the Max Planck Institute for Astrophysics, where this project was begun, to Guinevere Kauffmann for her advice and supervision during its earlier stages, and to Ann Zabludoff for her advice during the completion of this work; furthermore, to Antonaldo Diaferio, Huan Lin, Houjun Mo and the referee, Douglas Tucker, for valuable discussions and suggestions which shaped the contents of this paper.\\

\clearpage
\figcaption[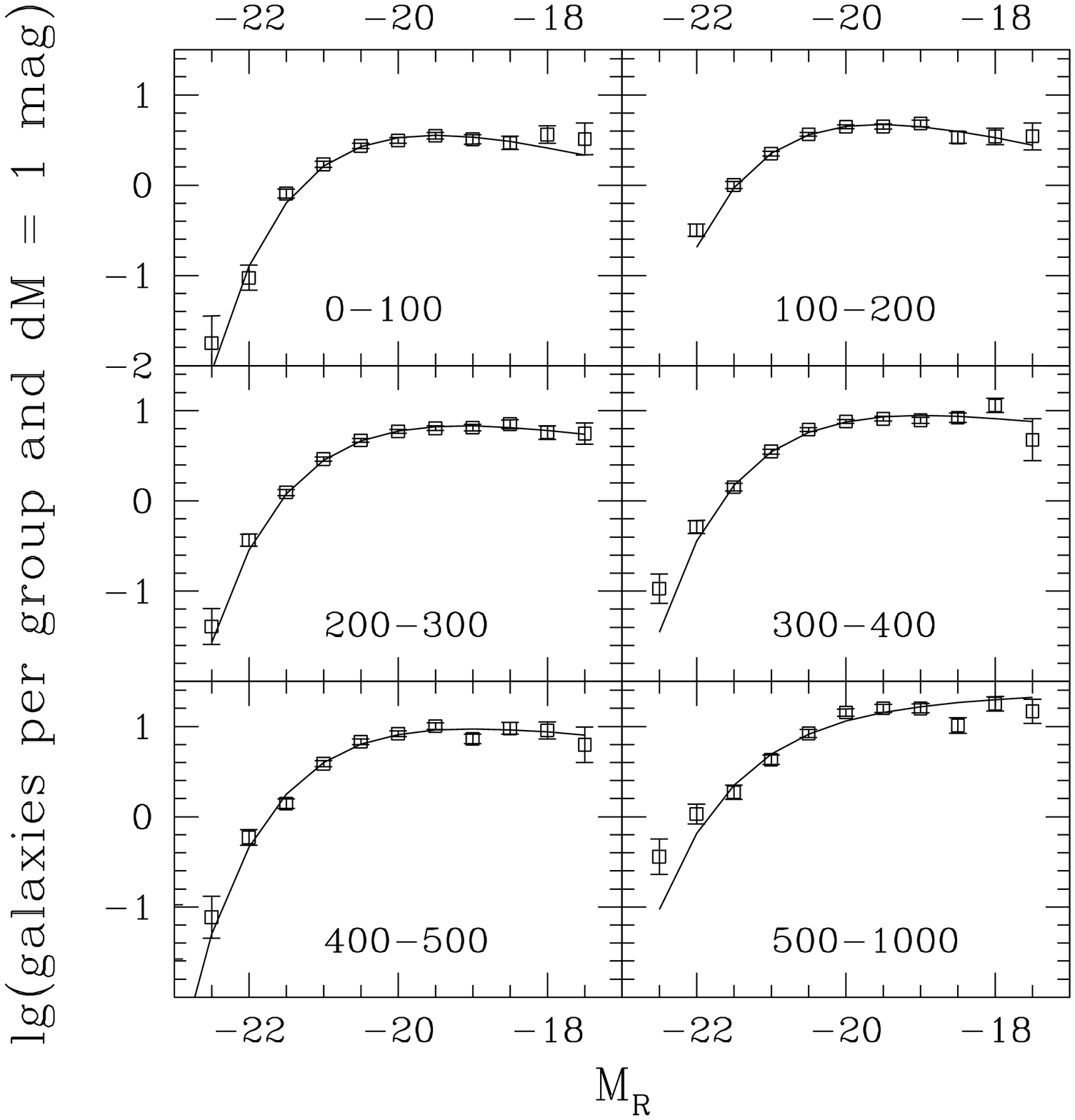]{Luminosity functions of group galaxies for various ranges of $\sigma$, independent of spectral type. Continuous lines are Schechter fits, data points discrete GLFs. Numbers give the range of $\sigma$ in km s$^{-1}$ for each subsample.}
\figcaption[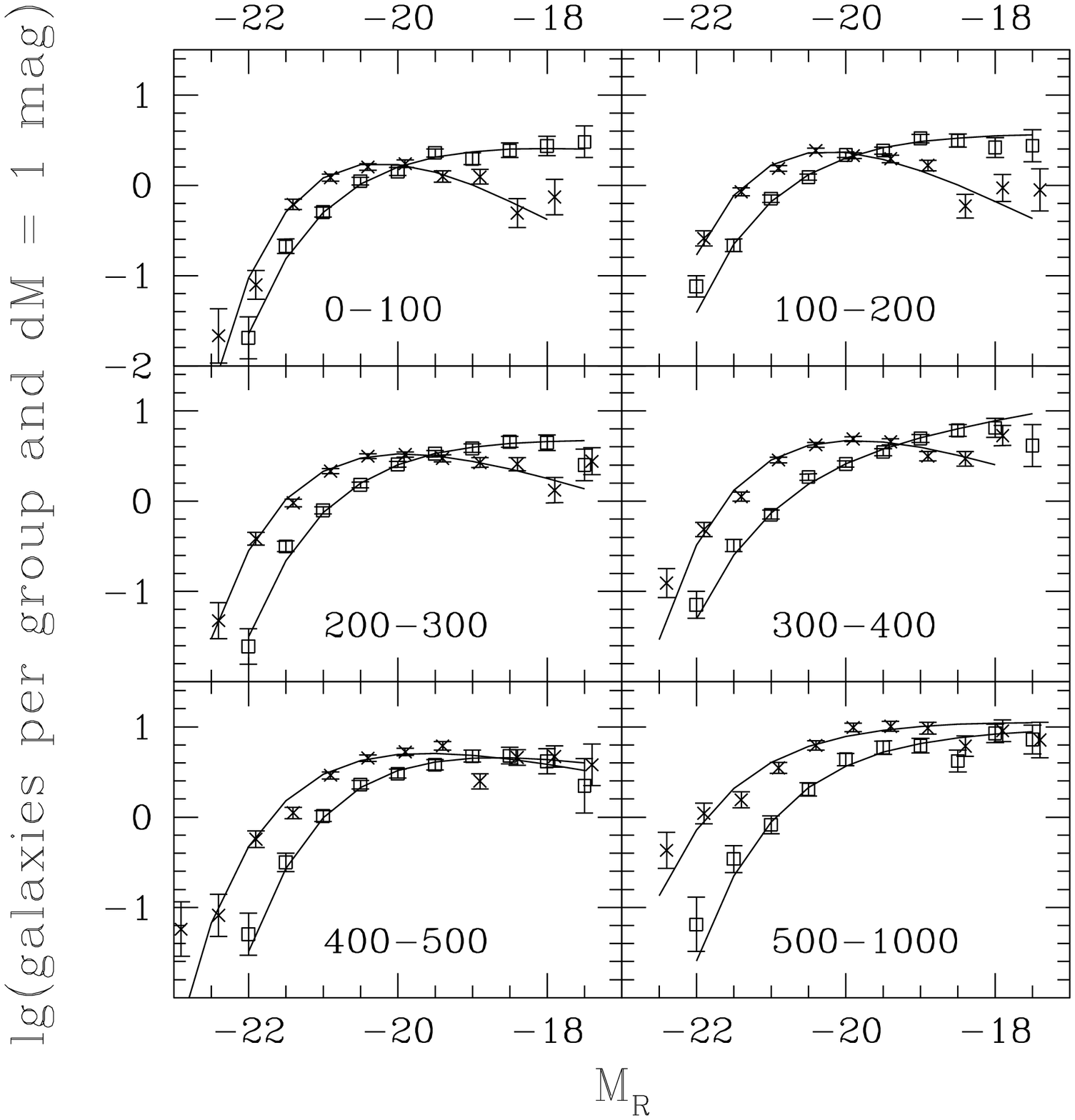]{Same as Fig. 1, but split up by spectral type. Crosses represent NEL, boxes EL galaxy subsamples.}
\figcaption[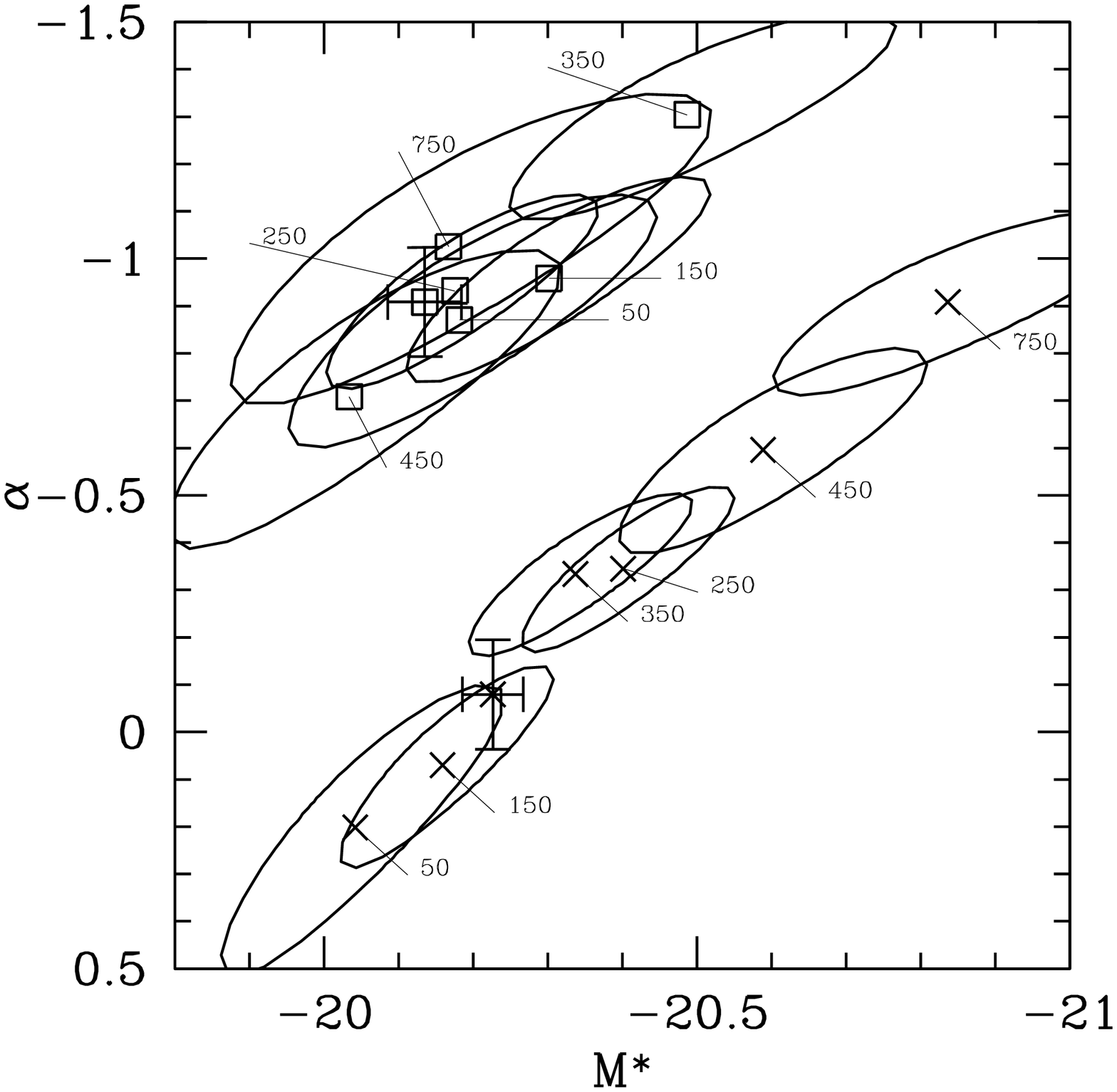]{$\alpha$-M$^{*}$ parameter space. Each subsample is represented by one data point. Squares represent EL subsamples, crosses NEL subsamples. Contour lines correspond to 1$\sigma$ uncertainties. Data points with error bars correspond to field galaxies. Numbers give the median $\sigma$ for each group subsample.}
\figcaption[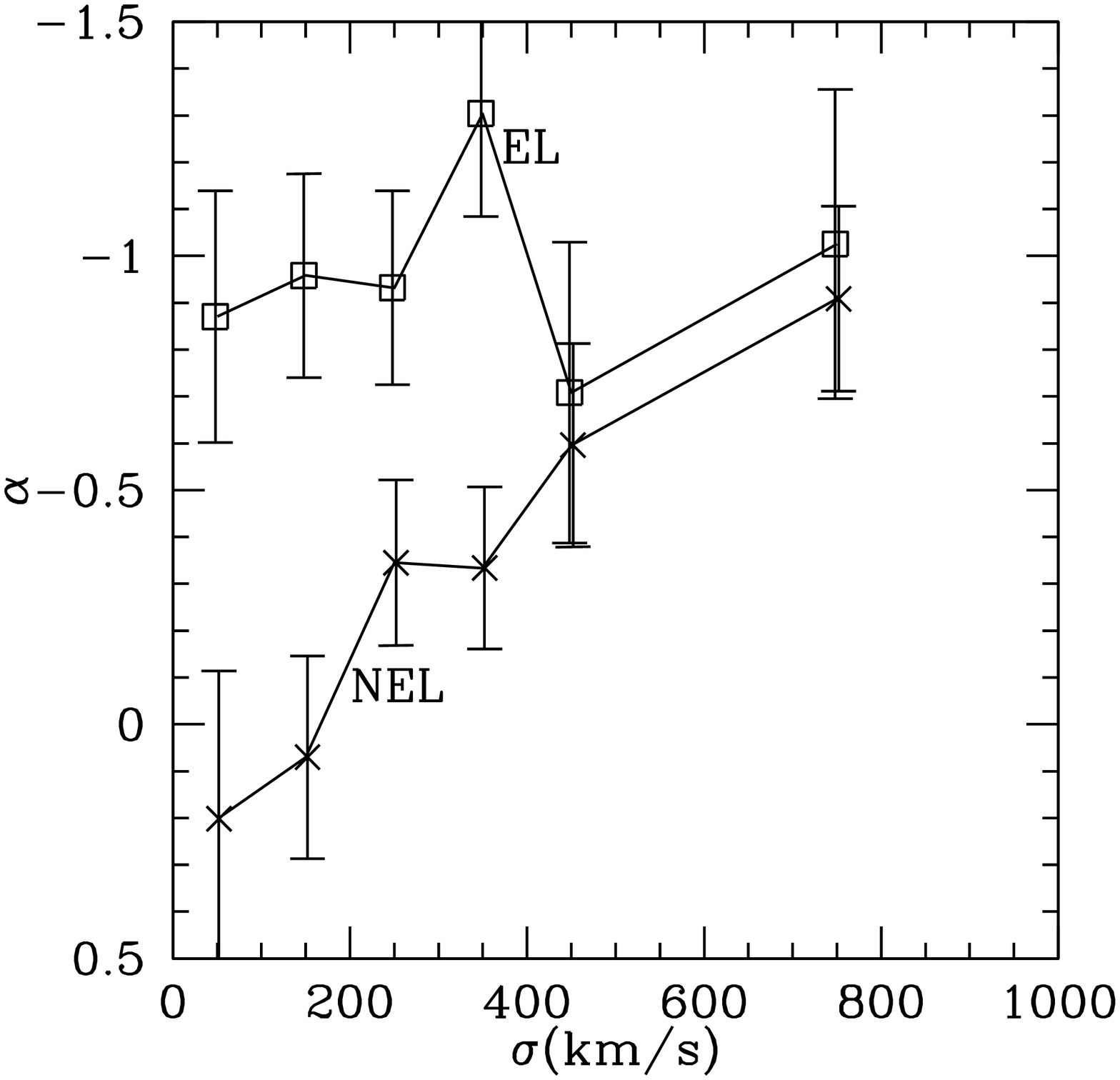]{$\alpha$ vs. $\sigma$. Error bars are 1$\sigma$. Symbols are the same as in Fig. 2}
\figcaption[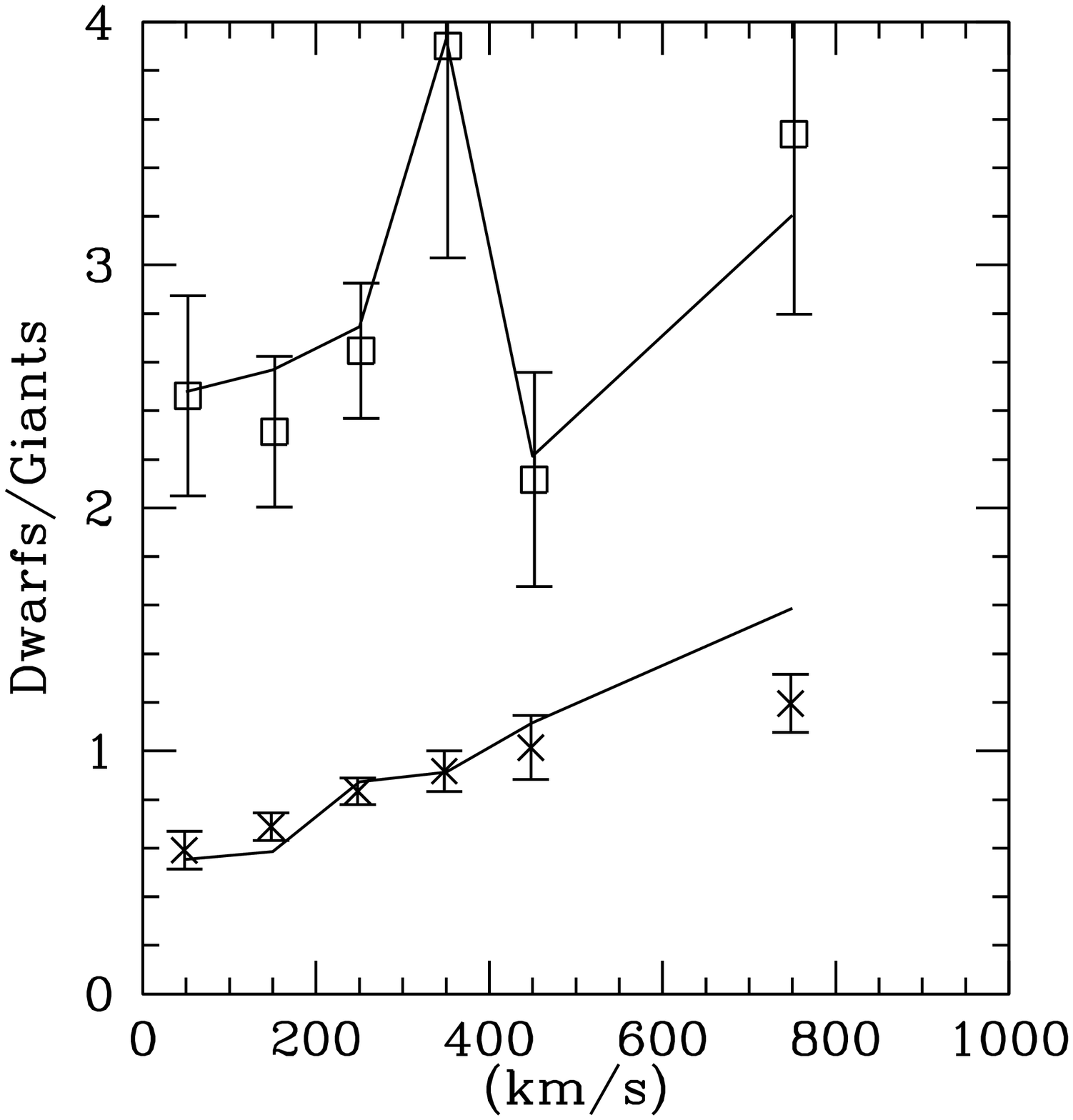]{Dwarf/giant ratio vs. $\sigma$. Symbols as in Fig. 2}
\figcaption[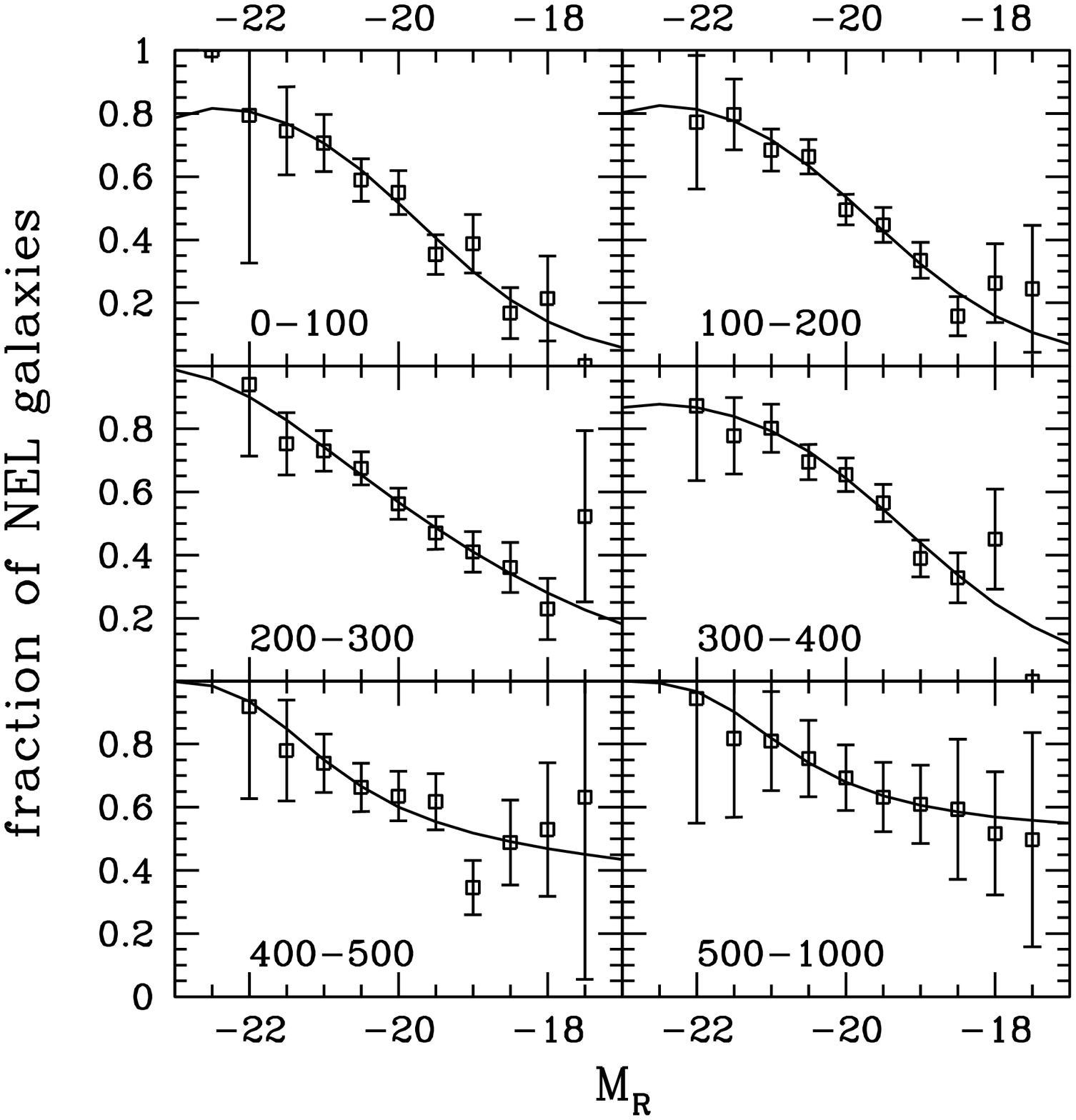]{Fraction of NEL galaxies as a function of M$_{R}$ for each $\sigma$ subsample. Continuous lines are derived from the parametric Schechter fits, while data points are derived from the discrete GLFs. Error bars are based on Poisson errors.}
\figcaption[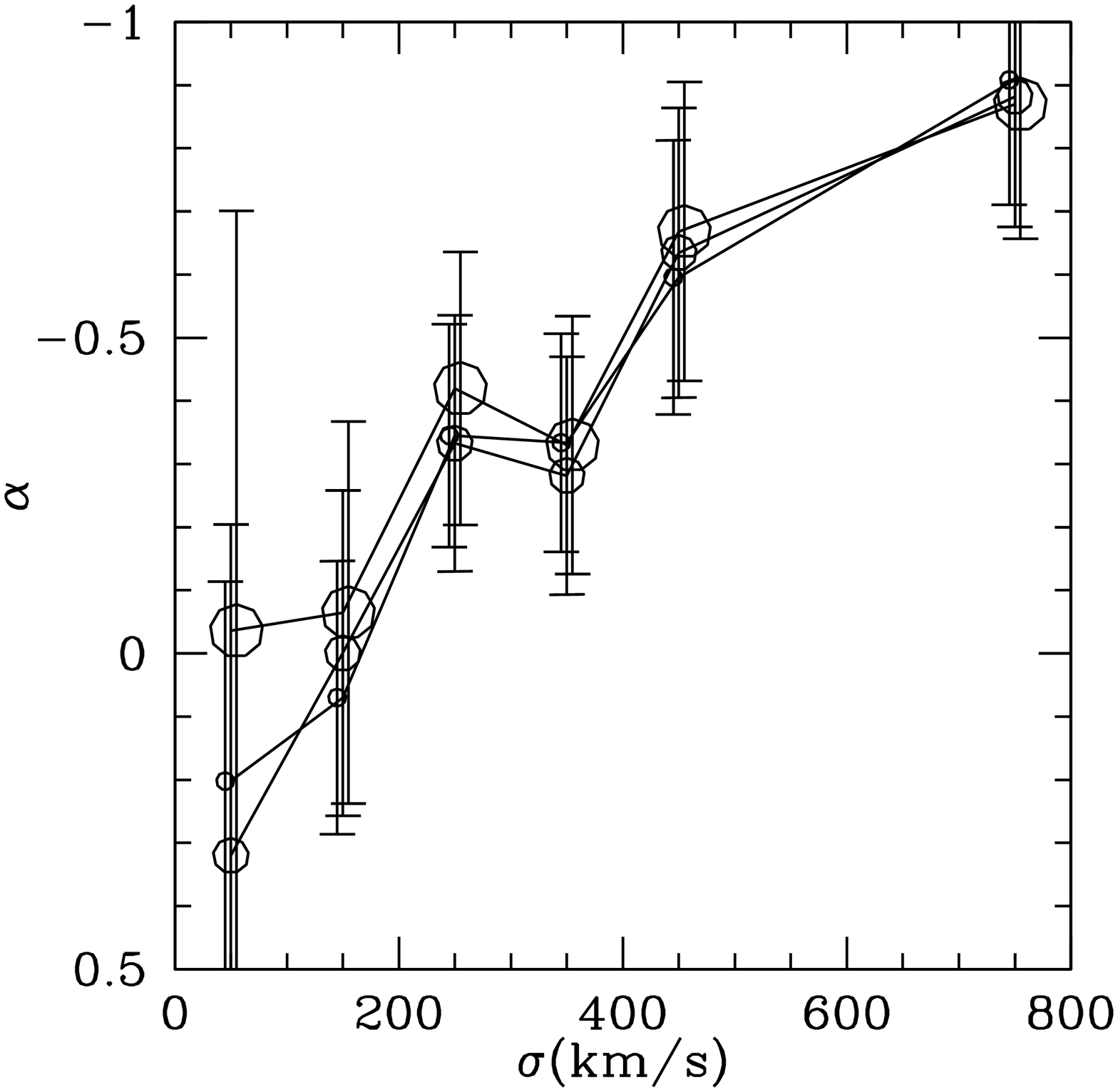] {$\alpha$ vs. $\sigma$ for NEL galaxies only for the full catalogue (small symbols), a catalogue with all galaxy triplets removed (medium-sized symbols) and a catalogue with triplets and quadruplets removed (large symbols)}
\figcaption[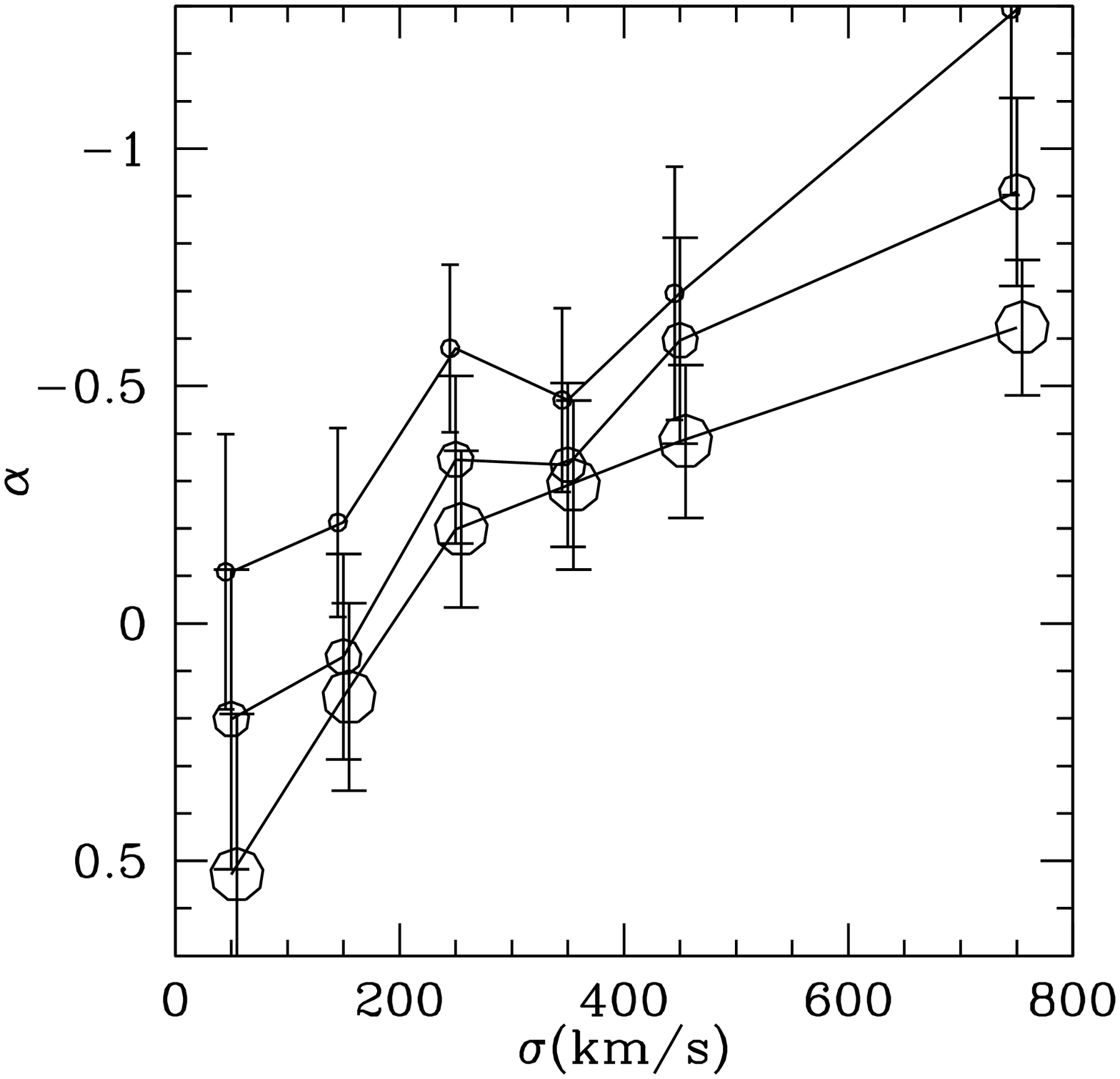]{$\alpha$ vs. $\sigma$ for NEL galaxies only for catalogues constructed with linking lengths of 0.715 Mpc / 500 km s$^{-1}$ (small symbols), 0.9 Mpc / 700 km s$^{-1}$ (medium-sized symbols) and 1.45 Mpc / 700 km s$^{-1}$ (large symbols)}
\figcaption[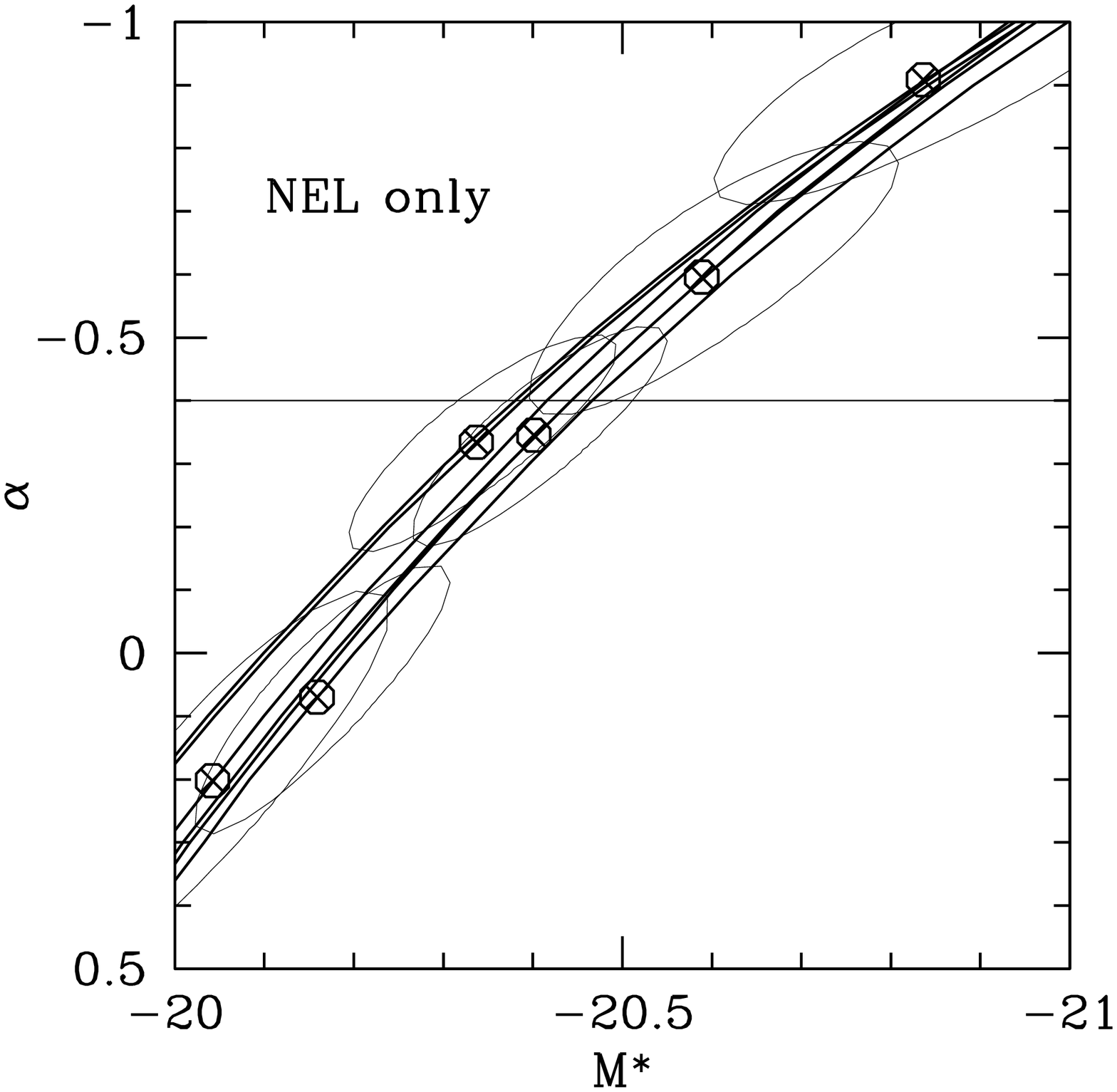]{Tracks in $\alpha$-M$^{*}$ space. The tracks show the best fit M$^{*}$ values for any given $\alpha$ for each subsample. Data points denote the best fit values for two degrees of freedom, as in Fig. 3. Error ellipses are 1$\sigma$. To evaluate the physical meaning of variations of M$^{*}$, subsamples should be compared for a given value of $\alpha$; $\alpha$ = -0.4 is highlighted as an example.}
\figcaption[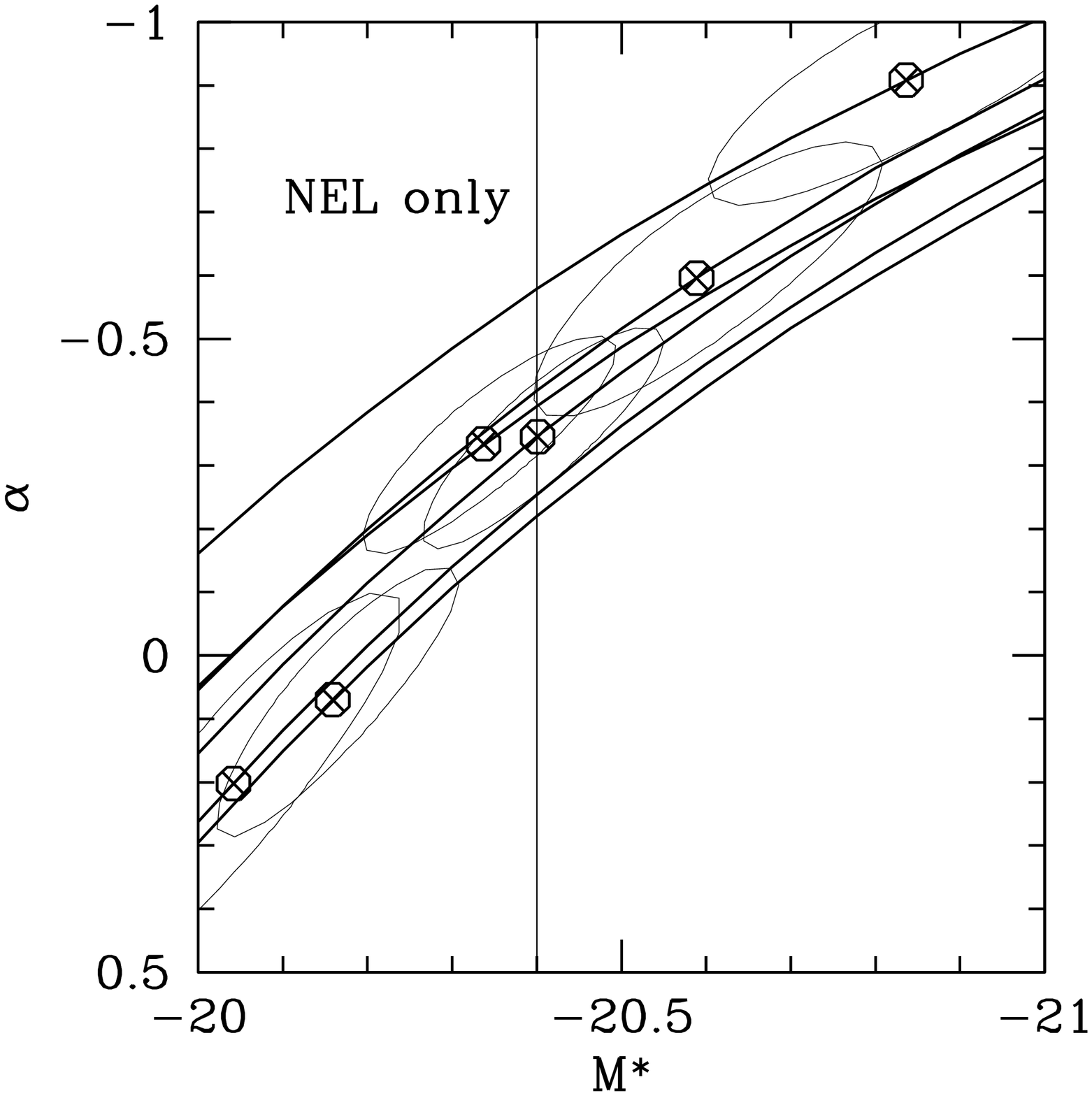]{Tracks in $\alpha$-M$^{*}$ space. The tracks show the best fit $\alpha$ values for any given M$^{*}$ for each subsample. Data points denote the best fit values for two degrees of freedom, as in Fig. 3. Error ellipses are 1$\sigma$. The vertical line highlights M$^{*}$ = -20.4 mag as an example of a given M$^{*}$ for which the $\alpha$ best fit values of the different samples may be compared.}

\clearpage
\begin{deluxetable}{lllll}
\tabletypesize{\scriptsize}
\tablecaption{Schechter parameters}
\tablewidth{0pt}
\tablehead{
\colhead{$\sigma$} & \colhead{$\alpha$} & \colhead{M*} & \colhead{$\Phi^{*}$} & \colhead{Galaxies} \\ \colhead{(km s$^{-1}$)} & & \colhead{(mag)} & \colhead {($group^{-1} mag^{-1}$)}
}
\startdata
combined samples\\
\tableline
field   &$-0.63\pm0.08$&$-20.29\pm0.03$&  ... &15802\\
$ 0-100$&$-0.49\pm0.19$&$-20.25\pm0.17$& 9.14 &867\\
$100-200$&$-0.50\pm0.15$&$-20.30\pm0.13$& 11.91 &1564\\
$200-300$&$-0.70\pm0.13$&$-20.42\pm0.12$& 14.36 &1846\\
$300-400$&$-0.74\pm0.13$&$-20.45\pm0.14$& 17.63 &1685\\
$400-500$&$-0.75\pm0.17$&$-20.50\pm0.18$& 18.68 &851\\
$500-1000$&$-1.01\pm0.17$&$-20.72\pm0.22$&23.83 &885\\
\tableline
NEL subsample\\
\tableline
field   &$-0.08\pm0.12$&$-20.23\pm0.04$&...&8280\\
$ 0-100$&$0.20\pm0.32$&$-20.04\pm0.20$&5.05&445\\
$100-200$&$0.07\pm0.22$&$-20.16\pm0.15$&6.94&842\\
$200-300$&$-0.35\pm0.18$&$-20.40\pm0.15$&9.19&1054\\
$300-400$&$-0.33\pm0.17$&$-20.34\pm0.16$&12.99&1048\\
$400-500$&$-0.60\pm0.22$&$-20.59\pm0.22$&11.93&532\\
$500-1000$&$-0.91\pm0.20$&$-20.84\pm0.27$&16.38&592\\
\tableline
EL subsample\\
\tableline
field   &$-0.91\pm0.12$&$-20.13\pm0.05$&...&7522\\
$ 0-100$&$-0.87\pm0.27$&$-20.18\pm0.26$&4.11&422\\
$100-200$&$-0.96\pm0.22$&$-20.30\pm0.22$&4.69&722\\
$200-300$&$-0.93\pm0.21$&$-20.18\pm0.19$&6.51&792\\
$300-400$&$-1.30\pm0.22$&$-20.49\pm0.28$&4.64&637\\
$400-500$&$-0.71\pm0.32$&$-20.03\pm0.28$&9.46&319\\
$500-1000$&$-1.03\pm0.33$&$-20.17\pm0.35$&11.03&293\\
\enddata
\end{deluxetable}

\clearpage
\begin{deluxetable}{lllll}
\tabletypesize{\scriptsize}
\tablecaption{$\alpha$-$\sigma$ correlation}
\tablewidth{0pt}
\tablehead{
\colhead{Subsample} & \colhead{Slope} & \colhead{Significance} & \colhead{$r_{S}$} \\      & \colhead{$(100 km s^{-1})^{-1}$} & \colhead{$\sigma$} & 
}
\startdata
combined & $-0.075\pm0.010$ & 8.3 & -1 \\
NEL	 & $-0.161\pm0.024$ & 7.7 & -0.94 \\
EL	 & $-0.010\pm0.045$ & 0.26 & -0.25 \\
\enddata
\end{deluxetable} 

\clearpage
\begin{deluxetable}{lllll}
\tabletypesize{\scriptsize}
\tablecaption{Dwarf/giant ratio}
\tablewidth{0pt}
\tablehead{
\colhead{Subsample} & \colhead{Slope} & \colhead{Significance} & \colhead{$r_{S}$} \\      & \colhead{$(100 km s^{-1})^{-1}$} & \colhead{$\sigma$} & 
}
\startdata
all, parametric & $(0.113 \pm 0.021) $ & 6.22 & 0.89 \\
all, discrete	& $(0.054 \pm 0.028) $ & 2.25 & 0.54 \\ 
NEL, parametric  & $(0.152 \pm 0.012) $ & 14.27 & 1 \\
NEL, discrete	& $(0.087 \pm 0.009) $ & 11.26 & 1 \\
EL, parametric & $(0.083 \pm 0.135) $ & 0.71 & 0.31 \\
EL, discrete & $(0.144 \pm 0.144) $ & 1.09 & 0.26 \\
\enddata
\end{deluxetable} 
  
\clearpage
\begin{deluxetable}{lllll}
\tabletypesize{\scriptsize}
\tablecaption{Mock catalogue results for $\alpha$}
\tablewidth{0pt}
\tablehead{
\colhead{Subsample} & \colhead{Slope} & \colhead{Significance} & \colhead{$r_{S}$} \\      & \colhead{$(100 km s^{-1})^{-1}$} & \colhead{$\sigma$} & 
}
\startdata
all & $(0.023 \pm 0.007) $ & 3.60 & 0.89 \\ 
EL  & $(0.005 \pm 0.017) $ & 0.29 & -0.18 \\
NEL & $(0.000 \pm 0.006) $ & 0.05 & 0.14 \\
\enddata
\end{deluxetable}

\clearpage
\begin{deluxetable}{lllll}
\tabletypesize{\scriptsize}
\tablecaption{Mock catalogue results for dwarf/giant ratio}
\tablewidth{0pt}
\tablehead{
\colhead{Subsample} & \colhead{Slope} & \colhead{Significance} & \colhead{$r_{S}$} \\      & \colhead{$(100 km s^{-1})^{-1}$} & \colhead{$\sigma$} & 
}
\startdata
all, parametric & $(-0.073 \pm 0.018) $ & 4.68 & -1 \\
all, discrete	& $(-0.060 \pm 0.021) $ & 3.36 & -0.89 \\ 
NEL, parametric  & $(-0.005 \pm 0.008) $ & 0.74 & -0.43 \\
NEL, discrete	& $(-0.005 \pm 0.012) $ & 0.44 & -0.43 \\
EL, parametric & $(-0.007 \pm 0.050) $ & 0.16 & -0.09 \\
EL, discrete & $(-0.006 \pm 0.057) $ & 0.13 & -0.14 \\
\enddata
\end{deluxetable}

\end {document}